 \documentclass[final,authoryear,3p,times,twocolumn]{elsarticle}

\usepackage[T1]{fontenc}
\usepackage{multicol}
\usepackage{setspace}
\usepackage{tocloft}
\usepackage{amsmath,bm}
\usepackage{listings}
\usepackage{caption}
\usepackage{lipsum}
\usepackage{graphicx}

\usepackage{verbatim}

\usepackage{multirow}
\usepackage{longtable}
\usepackage{float} 
\usepackage{booktabs}

\usepackage{fancyhdr}

\usepackage{amssymb}
\usepackage{float}
\restylefloat{figure}

\journal{Model of a synapse}

\begin{document}

\begin{frontmatter}



\title{Wave to pulse generation. From oscillatory synapse to train of action potentials}


\author{Alexandra Pinto Castellanos}
\address{Department of MNF UZH and D-ITET ETH UZH ETH}
\date{July 21, 2017 }

\begin{abstract}
Neurons have the capability of transforming information from a digital signal at the dendrites of the presynaptic terminal to an analogous wave at the synaptic cleft and back to a digital pulse when they achieve the required voltage for the generation of an action potential at the postsynaptic neuron. The main question of this research is what processes are generating the oscillatory wave signal at the synaptic cleft and what is the best model for this phenomenon. Here, it is proposed a model of the synapse as an oscillatory system capable of synchronization taking into account conservation of information and consequently of frequency at the interior of the synaptic cleft. Trains of action potentials certainly encode and transmit information along the nervous system but most of the time neurons are not transmitting action potentials, 99 percent of their time neurons are in the sub threshold regime were only small signals without the energy to emanate an action potential are carrying the majority of information. The proposed model for a synapse, smooths the train of action potential and keeps its frequency. Synapses are presented here as a system composed of an input wave that is transformed through interferometry. The collective synaptic interference pattern of waves will reflect the points of maximum amplitude for the density wave synaptic function were the location of the "particle" in our case action potential, has its highest probability. 
\end{abstract}

\end{frontmatter}


\section{Introducción}
\subsection{Synapses}
The structure of a neuron can be divided in three parts, receptor (dendritic tree), effector (axon)  and nucleus where the receptor and effector converge. The dendritic membrane forms synapses with the axon's tips of other neurons, with the special characteristic that there is no cytoplasmic bridge between them. The question at this point is: how information is conserved in this discontinuity? It is wildly known that dendrites receive input from hundreds of axon tips of other neurons, combine the input, which is delivered to the axon \cite{Purves}. However, how dendrites combine and deliver this information is not know with certainty, then is function of the axon to transmit the output of the dendrites to other parts of the nervous system.
\\ \\
The nature of information at the synaptic cleft and along the axons is very different in amplitude and time scale, however it is finely tuned in order to keep the flow and coherence of information. Dendrites from the postsynaptic neuron, receive an almost continuous oscillatory wave of ionic current input from the connecting synaptic cleft that joins it with the presynaptic terminal and convert it to a discrete voltage pulse that travels along the axon to be converted again in an oscillatory wave \cite{Bullock1993}. The wave to pulse basis transform takes place at the interior of each synapse, the mechanism is here proposed as follows: a pulse arrives on the axon terminal, this energy allows the entrance of positive calcium ions, which are going to move vesicles that carry neurotransmitters. This vesicles release their content outside the neuron where the oscillatory periodic wave takes the form of a field of ionic current. The mathematical details of this transformation are presented in the following sections. The output of the dendrites is the sum of waves resulting from all synaptic cleft activity, which are then delivered to the initial segment of the axon \cite{HODGKIN199025}, it is also presented a mathematical approach to this waves convergence.
\\ \\
There are two types of activity encoded at synapses, excitatory and inhibitory, depending on the type of neurotransmitter that is released in the synaptic cleft, this chemical wave if excite can let to synchronous behavior and generate a positive voltage or in case of inhibition can let to asynchronous behavior and generate a negative voltage in the postsynaptic neuron. It is proposed that at the synaptic cleft the signals are smoothed and convolved in order to create the oscillatory wave, that has the same frequency as the originating train but have the important property that can be synchronized with other synapses for amplification of information. This information is then transmitted in space and delayed at the initial segment of the axon \cite{Purves}. Once the wave has acquired enough amount of amplification after synchronization, the axon responds to this wave input by generating a pulse train, where each pulse has the same amplitude. But as a train of pulses, they keep the information flow frequency.

\subsection{Information flow}
The effect of all synapses working together, can excite or inhibit the production of a pulse in the receiver neuron. As mentioned before, this two effects are called excitatory synapses and inhibitory synapses, depending on the averaging effect of all the activity, the general output can be excitatory or inhibitory but not both. For the purposes of this paper, this general inhibitory or excitatory state, is the result of destructive or constructive wave pattern synchronization.
\\  \\
Until now, it is considered that if two or more small inputs are given simultaneously, the responses are simply added, and the system is said to be linear. If two otherwise identical inputs are separated in time, and if the responses are identical except for time of onset, the system is time invariant \cite{Eccles1964}.  Nonetheless, the mechanisms of synchronization, its variables, and limiting factors are unknown. On the other hand, it must be taken into account that the coupling medium is going to be crucial in the synchronization process, in the case of the synapse is mainly water, given that depending on the nonlinear nature of the synapse, it is possible to have in phase summation between presynaptic and postsynaptic neurons even when they are in anti phase, or it is possible to have anti phase behavior between the two cells even when they are in phase.
\\ \\
Another important effect that must be taken into account, is the secondary effect that the neurotransmitter release of a presynaptic terminal can have in neighbouring postsynaptic dendrites. There is a high possibility of having this cascade amplification given the analogous dispersive nature of the synapse signal and more importantly, given the required amplification of the signal.   
\\ \\
The wave to pulse conversion process, depends on the steady state level of the total dendritic synchronization, and its coincidence with the refractory period of the pulse (about 1 msec), during which no amount of stimuli can induce another pulse. Subsequently, there is a period lasting several milliseconds, in which an additional stimuli can induce a second pulse, only if the second stimuli is larger than the first. Given all this constraints, the wave to pulse transformation is nonlinear and variable in time, if a steady above-threshold current is passed across the membrane, the train pulse has a high frequency, but this frequency depend on many other parameters that make it variable in time.
\subsection{Cable Equation and chemical synapses}
The cable equation is a one-dimensional diffusion equation, and it has been widely used to model the dynamics of diffusion at chemical synapses. When a pulse discharges at the axon terminal, some of the vesicles in the pre-synaptic neuron discharge their contents into the synaptic cleft, this discharge has been understand as homogeneous. In the following sections I explain the oscillatory nature of this discharge, and in the characteristics that can be observed in the coupling between oscillators.
\\ \\
There is no biological proof of this proposed theory because the chemicals, ions and neurotransmitters directly in the synaptic cleft have not been measured. However, the concentration of this elements have been measured at the post-synaptic terminal, the concentration rises rapidly to a maximum value and then decays slowly\cite{Purves}, components of the chemical information can be detected many milliseconds after a maximum of activity.
\subsection{Synchronization of synapses}
Connectivity among neurons is required, and is present in massive amounts in order to generate synchronous coupling and amplification, with enough amplitude to transmit the analog oscillatory information of chemical synapses \cite{Hebb1949}. This activity should be nonlinear and self-sustaining, as was observed in biological oscillators. 
\\ \\
This nonlinear synchronization behavior was first studied in a bigger scale than in this thesis, and only taking into account the diffusion coupling by Katchalsky \cite{KATCHALSKY1976269}. He pointed out that a large number of nonlinear elements diffusely coupled, will give their properties of energy flow to the whole system. In this way, the system stabilises in a zero equilibrium state for E=0. Subsequently, the system is driven progressively away from equilibrium, until a critical level is reached where a phase transition occurs for $E\neq0$. 

\section{Steps of the theory}
Proposal and modelling of synapses as non linear oscillators capable of synchronization.  
\begin{itemize}
\item Research on the possibility and constraints for oscillation in the synapse.
\item Oscillator model of a synapse
\item Propose coupling between synapse
\item Synchronization constraints along synapses
\item Relation between spectrum of the total synaptic signal and the frequency spectrum of the action potential.
\end{itemize}

\subsection{Synchronization in the nervous System}
Biological oscillators like neurons and heart cells are usually modelled as nonlinear systems in a dimension 2 space. It is possible for this models to present limit cycles and synchronization. Nonlinear electronics as the Van Der Pol Oscillator and the Josephson Junction are good models for neuronal oscillators \citep{Strogatz}. \\ \\
A neuron has cycles of activity that oscillate between states of above threshold and sub threshold activity. Once the cell has the required energy, at a point in its cycle it releases an electrochemical signal that enables the communication channel to connect with other neurons in a synchronous fashion. Using the phase space of the neuron, let the description of how close the neuron is to reach the above threshold firing state. This cycle has a characteristic period and an amplitude. This means that the oscillations will remain near a constant frequency, even when the neuron is perturbed \citep{Matthews1991293}. 
\subsection{First theoretical study of Synchronization.}
Balthazar van Der Pol studied Synchronization theoretically for the first time. He worked synchronising triode generators from vacuum tubes. This research gave rise to the field of nonlinear dynamics given that he observed  that all initial conditions of the triode generator converged to the same final periodic orbit. In trying to model this behavior, they found equations of the phenomenon that couldn't be solved as we are used to solve linear equations, nonlinear Van Der Pol system of equations.
\begin{equation}
\ddot{X}+\mu \dot{X}(X^2-1)+X=0
\end{equation}
From the first and last term we can have an idea of the system that we are analyzing, and how we are going to model it because this two terms reflect the nature of an harmonic oscillator that can be achieved by the behaviour of an inductance and a capacitance. However, the term in the middle reflects a different oscillatory behavior, this is a damping nonlinear term that is originated by the functioning of vacuum tubes. We can see that the damping in the system can be positive or negative depending on the value of X. If $X>1$, it is a normal damping that tends to make the amplitude of the oscillations decay. But, if $X<1$ it is a negative damping which means pumping, this tends to amplify the amplitude of the oscillation. This pumping nonlinear effect is desirable for modelling the synapse given that neurons have a perpetual non zero activity that can be modelled through this term.
\section{Description of the biological phenomena}
The main interest of this research is to answer the wave to pulse generation problem in the chemical synapses of our nervous system. The current state of the art in this respect is pretty scarce and unclear, regarding conservation of information and frequency at the interior of the synaptic cleft. The curiosity to solve this problem was mainly raised by the fact that the trains of action potentials certainly encode and transmit information along the nervous system but most of the time neurons are not transmitting action potentials, 99 percent of their time they are in the sub threshold domain were only small signals without the energy to emanate an action potential are the ones that carry the majority of information, the one that let us perceive the world in one way, the same synchronised way that let us have a language, memory and in general, activities that do not require the fast response inter neuron communication as in electric synapses.\\ \\
The model that is proposed in this thesis for a synapse, is an oscillatory diffusion of ionic current, that smooths the train of action potential and keeps its frequency. Synapses can be seen as nonlinear, self-sustained oscillators with stable limit cycles, the linear oscillator, which can cycle at any amplitude, is not in agreement with real biological oscillators that have a regulated amplitude. The network of synapses should synchronise to generate the Gaussian properties of synchronous oscillators, this is achieved after the realisation of a Fourier series that transform from the analogous input of information that is given in each synapse oscillation to the almost discrete response along the axon in the form of an action potential. Synapses with the axons€™s tips of other neurons, with the special characteristic that there is no cytoplasmic bridge between them. The question at this point is: how information is conserved in this discontinuity?.\\ \\
The nature of information in dendrites and axons is completely opposite, yet the same. Dendrites receive an almost discrete pulse input and convert it to an analogous continuously oscillatory wave of ionic current\cite{Bullock1993}. The wave to pulse basis transform takes place at the interior of each synapse, the mechanism is as follows: a pulse arrives on the axon terminal, this energy allows the entrance of positive calcium ions, which are going to move vesicles that carry neurotransmitters. This vesicles release their content outside the neuron where the oscillatory periodic wave takes the form of a field of ionic current. The output of the dendrites is the sum of waves resulting from all pulse inputs, which is delivered to the initial segment of the axon \cite{Bullock1993} .
Locally, at the synapses, the signals are smoothed and convolved in order to
create the oscillatory wave, that has the same frequency as the train but can be
synchronized with other synapses for amplification of information. This informa-
tion is then transmitted in space and delayed at the initial segment of the axon
[Pur 2008]. Once the wave has acquired the enough amount of amplification after
synchronization, the axon responds to this wave input by generating a pulse train, 
where each pulse has the same amplitude. But as a train of pulses, they keep the
frequency information flow.\\ \\
The purpose of this thesis is the explanation and demonstration, that synapses
synchronization is necessary to generate a train of pulses along the axon. Addition-
ally, this synchronization is achieved when the spectrum of all the signal from the
synapses, matches the frequency spectrum of the action potential.
Presynapse as oscillator postsynapse as oscillator and synapse as medium that couple them to create
synthronization through the transmission of motion.
\\ \\
A neuron oscillates through oscillations in its
electrical field, and at a point in this cycle it releases an electrochemical signal to
connected neurons. The phase space of the neuron describes how close the neuron is
to firing this signal, and this cycle has both a characteristic period and an amplitude.
This means that the oscillations will remain near a constant frequency, even when the
neuron is perturbed.
\section{Theretical proposal}

Here I analyse the case in which a train of action potentials in the presynaptic neuron does not cause the postsynaptic one to fire, which is about 99\% of the neuron's life \cite{Katchalsky1976} and \cite{Freeman1975}. When a train of action potentials arrives at the synapse in the presynaptic neuron, it is converted into a chemical signal and this chemical signal in turn is converted into a small electrical signal that preserves the original frequency from the action potentials, see figure \ref{f:TrainNeurotransWave}.

 \begin{figure}[H]
    \centering
     \includegraphics[width=0.45\textwidth]{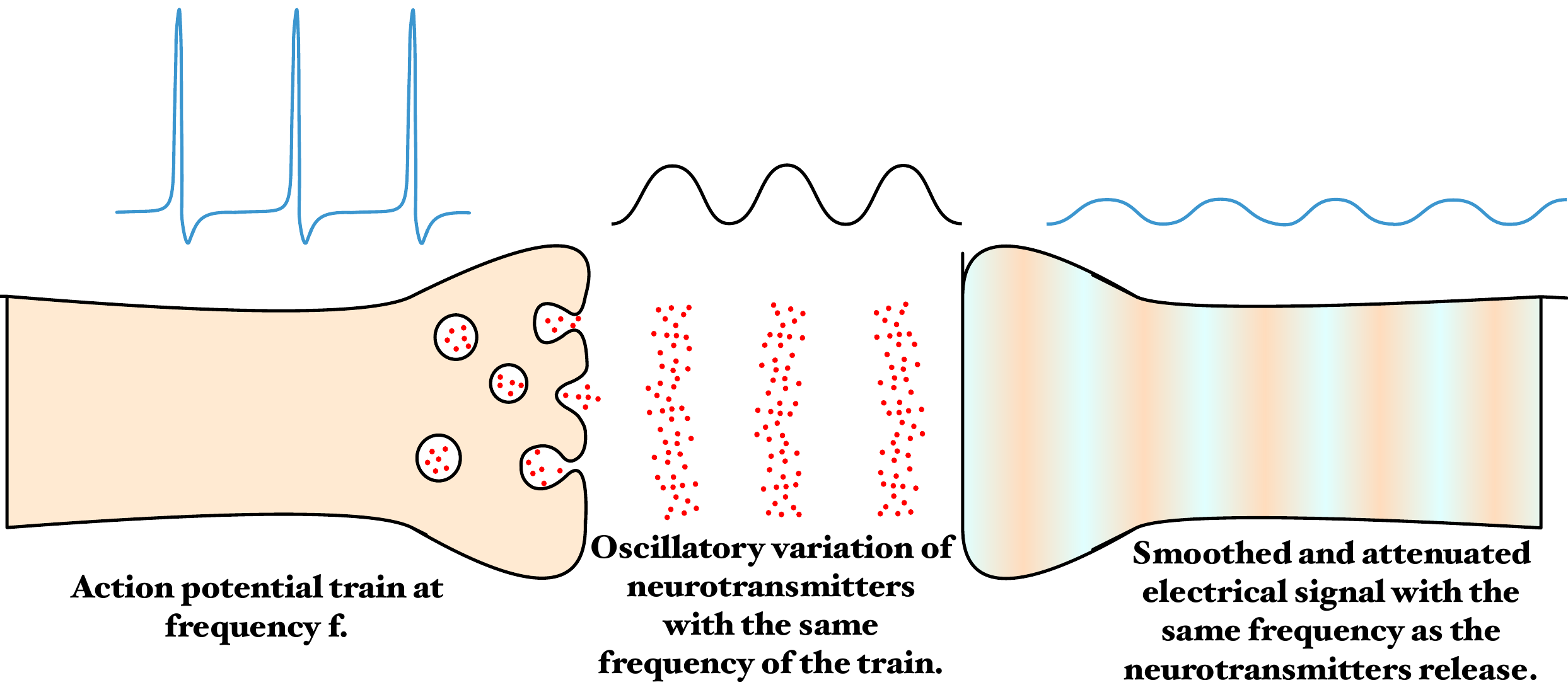}
     \caption{A train of action potentials in the presynaptic neuron is smoothed out by being converted into an oscillatory signal of neurotransmitter concentration in the synaptic cleft. This neurotransmitter concentration is then turned into a low amplitude oscillatory electrical signal in the postsynaptic neuron. The original frequency of the train of action potentials is preserved.}
    \label{f:sparse}
\end{figure}

\subsection{The synapse as a system}
\begin{figure}[H]
    \centering
     \includegraphics[width=0.45\textwidth]{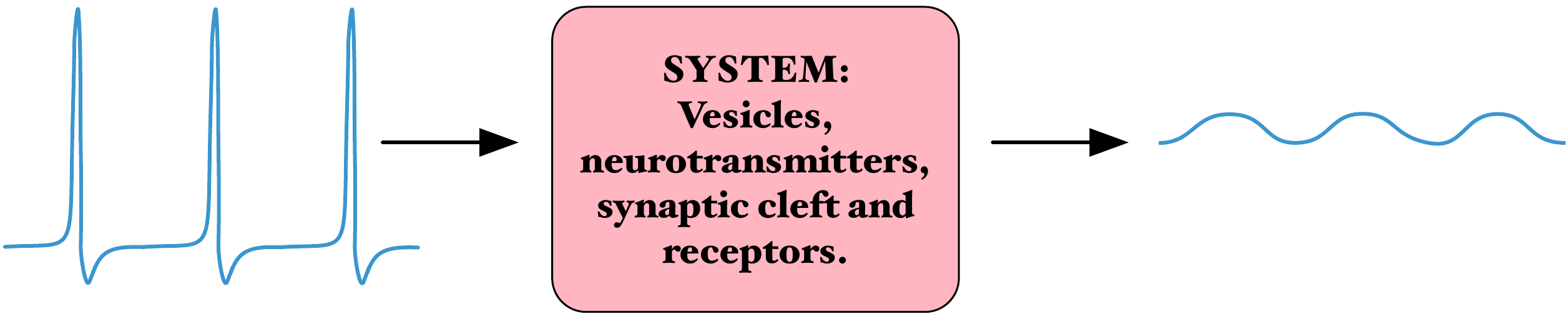}
     \caption{Vesicles, neurotransmitters and synaptic cleft as a system that transforms the train of spikes into a smoothed out version with the same primary frequency.}
    \label{f:SpikeSystemWave}
\end{figure}

When analyzing this process in more detail we observe that when the action potential reaches the tip of the axon of the presynaptic neuron, it triggers the movement of vesicles that release neurotransmitters that then travel through the synaptic cleft via a diffusive process towards the postsynaptic neuron. Afterwords an electrical signal is converted into a diffusive process modulated by the frequency of the pulse. Once the neurotransmitters reach the postsynaptic neuron, this oscillatory chemical signal is converted back into an oscillatory electrical signal. This process can be understood in the sense of chemical oscillations \cite{Turing1952}.\\ \\
This process can be modelled interpreting the train of action potentials as the input signal, and what happens with the vesicles, the synaptic cleft and the receptors can be interpreted as a system that transforms the input signal into a smoothed signal with lower amplitude. This idea can be seen in figure \ref{f:SpikeSystemWave}.

The following assumptions will apply for the model in figure \ref{f:SpikeSystemWave}: 
\begin{itemize}
\item The post-synaptic neuron will not emit an action potential, but will transmit small subthreshold signals.
\item The synaptic cleft dimensions stay the same, plasticity or changes in the configuration of the synapse are not happening.
\end{itemize}

If the above conditions are met the system in figure \ref{f:SpikeSystemWave} behaves as a time invariant system, which means that its output can be written as the input convolved with the impulse response of the system. In short words we can apply Fourier analysis to it. Thus, we can write:

\begin{equation}
\boxed{\textmd{Output}}=\boxed{\textmd{Input}}\,\ast\,\boxed{\textmd{Impulse Response of the System}}\label{e:MainIdeaAtTheCleft}
\end{equation}

\begin{figure}[H]
    \centering
     \includegraphics[width=0.45\textwidth]{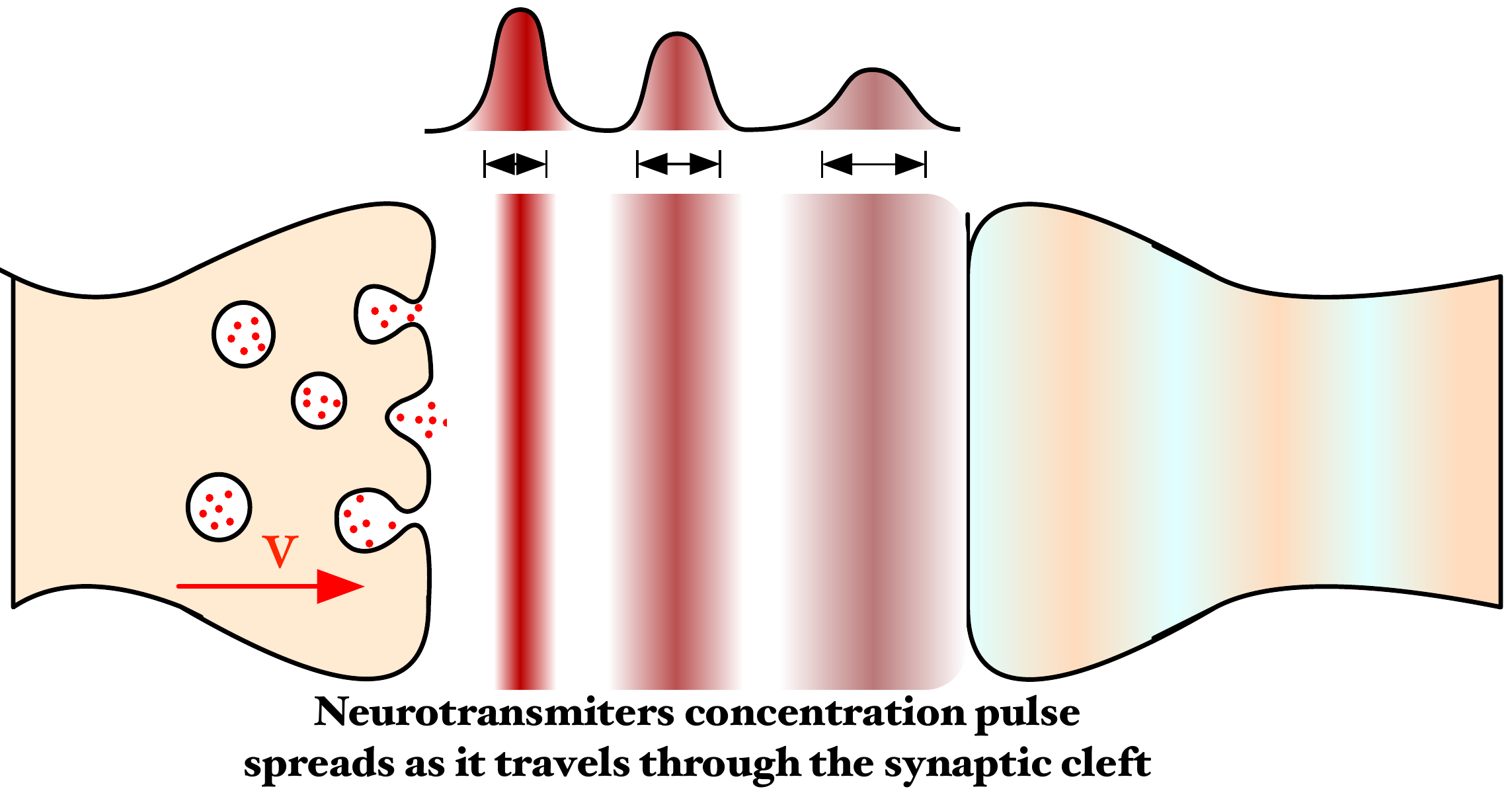}
     \caption{A neurotransmitter concentration leaves the presynaptic neuron with a certain velocity and as it travels through the synaptic cleft, it spreads by diffusion its concentration.}
    \label{f:SynapseDiffusion}
\end{figure}

To obtain the impulse response of the system, we need to analyse what happens at the synaptic cleft in more detail. When vesicles release neurotransmitters into the synaptic cleft, they release the concentration with approximately the same velocity at which the vesicles fused with the cellular membrane. As this blob of concentration travels through the synaptic cleft, it spreads by diffusion as seen in figure \ref{f:SynapseDiffusion}.

Therefore, similar to what has been described above, the postsynaptic neuron receives a smoothed concentration wave, which is then converted into a small periodic electrical signal, proportional to the original concentration.

\subsubsection{Vesicles and Diffusion at the synaptic cleft\label{ss:VesiclesDiffusionSynapCleft}}
The concentration gradient moves through the synaptic cleft, and as it moves it also diffuses, remembering that the medium is 80\% water at the interior of the cell. Thus, the postsynaptic neuron receives a diffused concentration potential. Assuming that the concentration pulse is on a fluid, moving at constant velocity for simplicity, the dominant effect on the spread of the signal is diffusion.\\

Diffusion in the direction along the synaptic cleft can be quantified, in the frame of the moving concentration pulse, by the 1-D diffusion equation:
\begin{equation}
\frac{\partial c(x,t)}{\partial t}=D\frac{\partial^2c(x,t)}{\partial x^2}\label{e:CDiffusion}
\end{equation}
This equation is solved applying the Fourier transform in space using:
\begin{equation*}
\mathcal{F}\lbrace g^{(n)}(x)\rbrace=(2\pi is)^n \mathcal{F}g(s)
\end{equation*}
 for the derivative. Thus equation \ref{e:CDiffusion} becomes:
\begin{equation}
\frac{\partial \mathcal{F}c(s,t)}{\partial t}=-2\pi^2s^2D\mathcal{F}c(s,t)\label{e:FourierDiffusion}
\end{equation}
Where $\mathcal{F}c(s,t)$, is the Fourier transform of the concentration. Equation \ref{e:FourierDiffusion} is a first order differential equation in time, therefore:
\begin{equation}
\mathcal{F}c(s,t)=\mathcal{F}c(s,0)e^{-2\pi^2s^2Dt}\label{e:TSolvedDiffusion}
\end{equation}
To solve this equation we use the following Fourier identities:
\begin{align*}
\frac{1}{\sqrt{2\pi t}}e^{-x^2/2t}& \stackrel{\mathrm{\mathcal{F}}}{\longleftrightarrow} e^{-2\pi^2s^2t}, & g(ax)&\stackrel{\mathrm{\mathcal{F}}}{\longleftrightarrow}\frac{1}{|a|}\mathcal{F}g\left(\frac{s}{a}\right), &\\
f\ast g&\stackrel{\mathrm{\mathcal{F}}}{\longleftrightarrow}(\mathcal{F}f)(\mathcal{F}g)
\end{align*}
Applying the inverse Fourier Transform to equation \ref{e:TSolvedDiffusion} we get:
\begin{align}
\mathcal{F}^{-1}\mathcal{F}c(s,t)&=\mathcal{F}^{-1}\mathcal{F}c(s,0)\ast\mathcal{F}^{-1}\lbrace e^{-2\pi^2s^2Dt}\rbrace  \nonumber \\
c(x,t)&=c(x,0)\ast \left[\frac{1}{\sqrt{2\pi Dt}}e^{\frac{-x^2}{2Dt}}\right]\label{e:SolvedDiffusion}
\end{align}
Equation \ref{e:SolvedDiffusion} prescribes how the concentration pulse spreads as it travels through the synaptic cleft. For now we assume that the synaptic cleft has a defined size and therefore the concentration pulse travels a predefined distance, which we will call $d_c$. Additionally the vesicles release the neurotransmitters at a velocity $v_n$ that in average is a constant. Therefore, the concentration pulse spreads over a time $t$ that we can predict:
\begin{align*}
t&=\frac{d_c}{v_n},      d_c=\textmd{ Synaptic cleft width}, \\ 
v_d&=\textmd{Velocity of neurotransmitters wave}
\end{align*}
The equations described in this model are a simplification of the main idea, that concentration pulses travel the synaptic cleft at a more or less standard time, since most synaptic clefts exhibit the same dimensions. Thus we can write the total spread, that the concentration pulse acquires while traveling through the synaptic cleft as follows:
\begin{equation}
c(x,d_c/v_n)=c(x,0)\ast \left[\frac{1}{\sqrt{2\pi D(d_c/v_n)}}e^{\frac{-x^2}{2D(d_c/v_n)}}\right]\label{e:ConcentrationSpread}
\end{equation}
Additionally, we can now rename the diffusion constant $D$, $d_c$ and $v_n$ by one single constant, $\zeta$, that quantifies the spread of the signal as it travels through the synaptic cleft, as shown in the following equation:
\begin{equation}
c(x,d_c/v_n)=c(x,0)\ast \left[\frac{1}{\sqrt{2\pi \zeta}}e^{\frac{-x^2}{2\zeta}}\right]\label{e:ConcentrationSpreadFinal}
\end{equation}
We have to remember that this solution is derived in the coordinate system of the pulse where the concentration pulse is emitted by the presynaptic neuron and is absorbed by the postsynaptic neuron. 
Furthermore, the concentration of neurotransmitters will be a function of the voltage in the neuron where, $c(t)=Ap(t)$ and with this in mind, we can rewrite the concentration pulse in the neuron as a voltage function of time for both the pre and postsynaptic neuron: The following equations describe the following sequence of events 1- Voltage converted to concentration pulse. Pre-synaptic. 2- Concentration converted to voltage. Post-synaptic. 3- Spatial convolution converted to time convolution at the synapse.

\begin{align}
c(x,0)&=A_{in}p_{in}(t), \\ 
c(x,d_c/v_n)&=A_{out}p_{out}(t), \\
e^{\frac{-x^2}{2\zeta}} &= e^{\frac{-(A_{in}t)^2}{2\zeta}}, 
\end{align}
As mentioned before, I want to reiterate that the neuron converts voltage signals to concentration signals at the presynaptic neuron and concentration signals to voltage at the postsynaptic neuron. This happens in a similar way, as a video camera in our phone converts a visual signal into a digital one, and then the screen converts a digital signal into a visual signal. Based on this analogy, the spatial convolution in equation \ref{e:ConcentrationSpreadFinal} can be converted into a time convolution:
\begin{equation}
A_{out}p_{out}(t)=A_{in}p_{in}(t) \ast \left[\frac{1}{\sqrt{2\pi \zeta}}e^{\frac{-(A_{in}t)^2}{2\zeta}}\right]\label{e:DiffusionToTime}
\end{equation}
We have differentiated the voltage-to-concentration and concentration-to-voltage conversions at the pre and post-synaptic neuron respectively, because these two types of conversions are governed by different underlying processes. Rearranging equation \ref{e:DiffusionToTime}, we have:
\begin{equation}
\boxed{p_{out}(t)=\frac{1}{A_{out}}p_{in}(t) \ast \left[\frac{1}{\sqrt{2\pi \zeta/A_{in}^2}}e^{\frac{-t^2}{2\zeta/A_{in}^2}}\right]}\label{e:CleftFinal}
\end{equation}

\begin{figure}[H]
    \centering
     \includegraphics[width=0.45\textwidth]{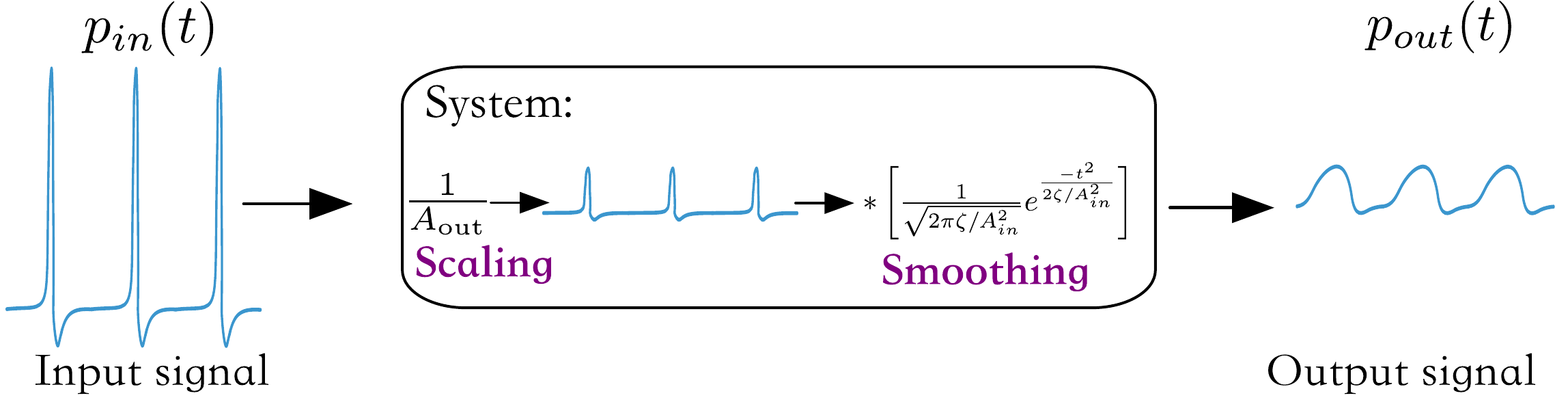}
     \caption{The synapse as a energy conversion and smoothing system.}
    \label{f:SpikesConvolutionWave}
\end{figure}

Equation \ref{e:CleftFinal} explains what happens to the action potential as it reaches the end of the presynaptic neuron, which is converted to a chemical signal and then back into an electrical signal. The scaling constant $1/A_{out}$ quantifies how much the amplitude of the signal is reduced. Additionally, the Gaussian function present in equation \ref{e:CleftFinal} is normalized, thus the convolution smooths out the signal. We now have a mathematical representation of the system in figure \ref{f:SpikeSystemWave}. To understand better what happens in equation \ref{f:SpikeSystemWave}, we can appeal to figure \ref{f:SpikesConvolutionWave}, where the input signal of spikes or pulses is smoothed and becomes small. The system box in figure \ref{f:SpikesConvolutionWave} depicts the behaviour of the synapse.

We now have a mathematical explanation for the system in figure \ref{f:SpikeSystemWave} and the conceptual equation \ref{e:MainIdeaAtTheCleft}, where we have explained the vesicles, neurotransmitters, synaptic cleft, and receptors as a system quantified in our final result in equation \ref{e:CleftFinal}.

\subsubsection{Predictive model for a train of action potentials crossing the synaptic cleft just after conversion into neurotransmitters.\label{ss:TrainActionCrossingSyn}}
In the previous section we have just derived what happens at the synapse and now we apply the result of equation \ref{e:CleftFinal}, to predict what will happen to a train of action potentials.\\

\begin{figure}[H]
    \centering
     \includegraphics[width=0.45\textwidth]{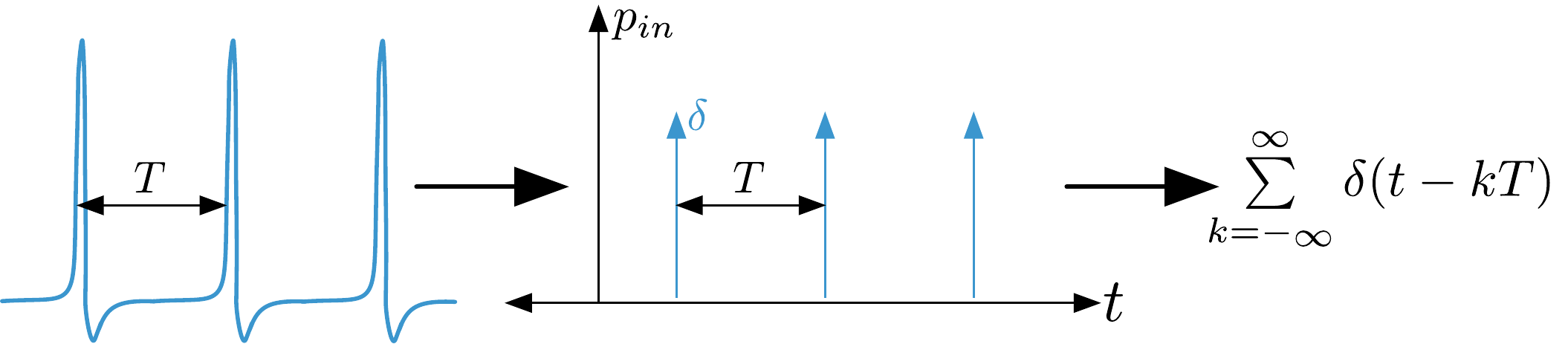}
     \caption{A train of action potentials written as Dirac's delta functions. The time interval between two action potentials is $T$.}
    \label{f:Deltitas}
\end{figure}

First we need to write a train of action potentials in mathematical terms and for now I propose to model them as a train of Dirac's delta functions, as this captures their main characteristics, especially the fact that they arrive at a certain frequency at the end of the synaptic neuron. This idea is illustrated in figure \ref{f:Deltitas}.

We can express the train of action potentials as a train of Dirac's delta functions yet, in reality we will have a bounded train of action potentials. However, for now and without loss of generality, I will write it as an infinite train of deltas as it is a well known function in Fourier analysis:
\begin{equation}
\textmd{Action potentials spaced T}=\sum\limits_{k=-\infty}^\infty\delta(t-kT)=\xi_T(t)\label{e:Sha}
\end{equation}
With this formalism for the action potentials, $\xi_T(t)$, we can now see what will happen to the electric signal as it travels through the synapse. Using our result from equation \ref{e:CleftFinal} and equation \ref{e:Sha}, we have:
\begin{equation}
p_{in}(t)=\xi_T(t)
\end{equation}
Thus:
\begin{align}
p_{out}(t)&=\frac{1}{A_{out}}\xi_T(t) \ast \left[\frac{1}{\sqrt{2\pi \zeta/A_{in}^2}}e^{\frac{-t^2}{2\zeta/A_{in}^2}}\right]\\[1ex]
&=\frac{1}{A_{out}}\left[\sum\limits_{k=-\infty}^\infty\delta(t-kT)\right] \ast \left[\frac{1}{\sqrt{2\pi \zeta/A_{in}^2}}e^{\frac{-t^2}{2\zeta/A_{in}^2}}\right]\\
&=\frac{1}{A_{out}}\sum\limits_{k=-\infty}^\infty\delta(t-kT) \ast \frac{1}{\sqrt{2\pi \zeta/A_{in}^2}}e^{\frac{-t^2}{2\zeta/A_{in}^2}}
\end{align}
And using the fact that $\delta(t-a)\ast f(t)=f(t-a)$, we obtain:
\begin{equation}
\boxed{p_{out}(t)=\frac{1}{A_{out}}\frac{1}{\sqrt{2\pi \zeta/A_{in}^2}}e^{\frac{-(t-kT)^2}{2\zeta/A_{in}^2}}}\label{e:GaussWave}
\end{equation}
This is a surprising result, since a train of action potentials has become a train of smooth Gaussian functions, which perfectly resembles a wave. Now, what we quantitatively saw in figure \ref{f:SpikesConvolutionWave} is expressed in equation \ref{e:GaussWave}. The synapse converts a train of action potentials into a smoothed oscillatory signal. From all of this follows that the behaviour of a subthreshold signal going into a postsynaptic neuron can easily be modeled by an oscillator that outputs a Gaussian like wave. Furthermore, this behaviour has been studied by expressing a train of action potentials as deltas \cite{Kandel1961}. When we represent an action potential in more detail, the signal will be even closer to a wave and since a delta function is the sharpest signal that can exist, it can be turned into a smoother gaussian. Therefore, a train of action potentials, which is a continuous signal, will be more smoothed out by the synapse and the output will become closer to a sinusoidal wave.\\

With this, I have demonstrated that for the purpose of studying the signals going into a single neuron, we can substitute the input synapse by an oscillator that outputs waves of a predefined frequency. 
\section{Discussion}
Even though there is no biological proof regarding the synapse as an oscillator, there are many advantages for neural networks applications if we model the synapse as an oscillator. If we consider a detailed biophysical model of the synapse and we analyse the behavior of the different components, from calcium dynamics to vesicles release, it is possible to see the oscillatory activity. Those biophysical models make use of experimental data in order to solve the coupling of differential equations in a numerical way through the tuning of variables. In this way, we can be more confident about the assumption of oscillations along the presynaptic terminal, synaptic cleft and postsinaptic receptors. \\ \\
It is through wave interference patterns that this model generates the enough amount of fluctuations and coherence to represent the activity recorded in the postsynapse.

\nocite{*}
\bibliographystyle{elsarticle-harv}
\bibliography{Bib}






\end{document}